\documentclass[a4paper]{jpconf}
\usepackage{graphicx,hyperref}
\usepackage{todonotes}
\usepackage{algorithm2e}
\usepackage{algorithmic}
\algsetup{linenosize=\tiny}
\usepackage[draft]{minted}
\usepackage[labelfont=bf]{caption}

\date{\today}

\makeatletter
\def\PYG@reset{\let\PYG@it=\relax \let\PYG@bf=\relax%
    \let\PYG@ul=\relax \let\PYG@tc=\relax%
    \let\PYG@bc=\relax \let\PYG@ff=\relax}
\def\PYG@tok#1{\csname PYG@tok@#1\endcsname}
\def\PYG@toks#1+{\ifx\relax#1\empty\else%
    \PYG@tok{#1}\expandafter\PYG@toks\fi}
\def\PYG@do#1{\PYG@bc{\PYG@tc{\PYG@ul{%
    \PYG@it{\PYG@bf{\PYG@ff{#1}}}}}}}
\def\PYG#1#2{\PYG@reset\PYG@toks#1+\relax+\PYG@do{#2}}

\expandafter\def\csname PYG@tok@gd\endcsname{\def\PYG@tc##1{\textcolor[rgb]{0.63,0.00,0.00}{##1}}}
\expandafter\def\csname PYG@tok@gu\endcsname{\let\PYG@bf=\textbf\def\PYG@tc##1{\textcolor[rgb]{0.50,0.00,0.50}{##1}}}
\expandafter\def\csname PYG@tok@gt\endcsname{\def\PYG@tc##1{\textcolor[rgb]{0.00,0.27,0.87}{##1}}}
\expandafter\def\csname PYG@tok@gs\endcsname{\let\PYG@bf=\textbf}
\expandafter\def\csname PYG@tok@gr\endcsname{\def\PYG@tc##1{\textcolor[rgb]{1.00,0.00,0.00}{##1}}}
\expandafter\def\csname PYG@tok@cm\endcsname{\let\PYG@it=\textit\def\PYG@tc##1{\textcolor[rgb]{0.25,0.50,0.50}{##1}}}
\expandafter\def\csname PYG@tok@vg\endcsname{\def\PYG@tc##1{\textcolor[rgb]{0.10,0.09,0.49}{##1}}}
\expandafter\def\csname PYG@tok@vi\endcsname{\def\PYG@tc##1{\textcolor[rgb]{0.10,0.09,0.49}{##1}}}
\expandafter\def\csname PYG@tok@mh\endcsname{\def\PYG@tc##1{\textcolor[rgb]{0.40,0.40,0.40}{##1}}}
\expandafter\def\csname PYG@tok@cs\endcsname{\let\PYG@it=\textit\def\PYG@tc##1{\textcolor[rgb]{0.25,0.50,0.50}{##1}}}
\expandafter\def\csname PYG@tok@ge\endcsname{\let\PYG@it=\textit}
\expandafter\def\csname PYG@tok@vc\endcsname{\def\PYG@tc##1{\textcolor[rgb]{0.10,0.09,0.49}{##1}}}
\expandafter\def\csname PYG@tok@il\endcsname{\def\PYG@tc##1{\textcolor[rgb]{0.40,0.40,0.40}{##1}}}
\expandafter\def\csname PYG@tok@go\endcsname{\def\PYG@tc##1{\textcolor[rgb]{0.53,0.53,0.53}{##1}}}
\expandafter\def\csname PYG@tok@cp\endcsname{\def\PYG@tc##1{\textcolor[rgb]{0.74,0.48,0.00}{##1}}}
\expandafter\def\csname PYG@tok@gi\endcsname{\def\PYG@tc##1{\textcolor[rgb]{0.00,0.63,0.00}{##1}}}
\expandafter\def\csname PYG@tok@gh\endcsname{\let\PYG@bf=\textbf\def\PYG@tc##1{\textcolor[rgb]{0.00,0.00,0.50}{##1}}}
\expandafter\def\csname PYG@tok@ni\endcsname{\let\PYG@bf=\textbf\def\PYG@tc##1{\textcolor[rgb]{0.60,0.60,0.60}{##1}}}
\expandafter\def\csname PYG@tok@nl\endcsname{\def\PYG@tc##1{\textcolor[rgb]{0.63,0.63,0.00}{##1}}}
\expandafter\def\csname PYG@tok@nn\endcsname{\let\PYG@bf=\textbf\def\PYG@tc##1{\textcolor[rgb]{0.00,0.00,1.00}{##1}}}
\expandafter\def\csname PYG@tok@no\endcsname{\def\PYG@tc##1{\textcolor[rgb]{0.53,0.00,0.00}{##1}}}
\expandafter\def\csname PYG@tok@na\endcsname{\def\PYG@tc##1{\textcolor[rgb]{0.49,0.56,0.16}{##1}}}
\expandafter\def\csname PYG@tok@nb\endcsname{\def\PYG@tc##1{\textcolor[rgb]{0.00,0.50,0.00}{##1}}}
\expandafter\def\csname PYG@tok@nc\endcsname{\let\PYG@bf=\textbf\def\PYG@tc##1{\textcolor[rgb]{0.00,0.00,1.00}{##1}}}
\expandafter\def\csname PYG@tok@nd\endcsname{\def\PYG@tc##1{\textcolor[rgb]{0.67,0.13,1.00}{##1}}}
\expandafter\def\csname PYG@tok@ne\endcsname{\let\PYG@bf=\textbf\def\PYG@tc##1{\textcolor[rgb]{0.82,0.25,0.23}{##1}}}
\expandafter\def\csname PYG@tok@nf\endcsname{\def\PYG@tc##1{\textcolor[rgb]{0.00,0.00,1.00}{##1}}}
\expandafter\def\csname PYG@tok@si\endcsname{\let\PYG@bf=\textbf\def\PYG@tc##1{\textcolor[rgb]{0.73,0.40,0.53}{##1}}}
\expandafter\def\csname PYG@tok@s2\endcsname{\def\PYG@tc##1{\textcolor[rgb]{0.73,0.13,0.13}{##1}}}
\expandafter\def\csname PYG@tok@nt\endcsname{\let\PYG@bf=\textbf\def\PYG@tc##1{\textcolor[rgb]{0.00,0.50,0.00}{##1}}}
\expandafter\def\csname PYG@tok@nv\endcsname{\def\PYG@tc##1{\textcolor[rgb]{0.10,0.09,0.49}{##1}}}
\expandafter\def\csname PYG@tok@s1\endcsname{\def\PYG@tc##1{\textcolor[rgb]{0.73,0.13,0.13}{##1}}}
\expandafter\def\csname PYG@tok@ch\endcsname{\let\PYG@it=\textit\def\PYG@tc##1{\textcolor[rgb]{0.25,0.50,0.50}{##1}}}
\expandafter\def\csname PYG@tok@m\endcsname{\def\PYG@tc##1{\textcolor[rgb]{0.40,0.40,0.40}{##1}}}
\expandafter\def\csname PYG@tok@gp\endcsname{\let\PYG@bf=\textbf\def\PYG@tc##1{\textcolor[rgb]{0.00,0.00,0.50}{##1}}}
\expandafter\def\csname PYG@tok@sh\endcsname{\def\PYG@tc##1{\textcolor[rgb]{0.73,0.13,0.13}{##1}}}
\expandafter\def\csname PYG@tok@ow\endcsname{\let\PYG@bf=\textbf\def\PYG@tc##1{\textcolor[rgb]{0.67,0.13,1.00}{##1}}}
\expandafter\def\csname PYG@tok@sx\endcsname{\def\PYG@tc##1{\textcolor[rgb]{0.00,0.50,0.00}{##1}}}
\expandafter\def\csname PYG@tok@bp\endcsname{\def\PYG@tc##1{\textcolor[rgb]{0.00,0.50,0.00}{##1}}}
\expandafter\def\csname PYG@tok@c1\endcsname{\let\PYG@it=\textit\def\PYG@tc##1{\textcolor[rgb]{0.25,0.50,0.50}{##1}}}
\expandafter\def\csname PYG@tok@o\endcsname{\def\PYG@tc##1{\textcolor[rgb]{0.40,0.40,0.40}{##1}}}
\expandafter\def\csname PYG@tok@kc\endcsname{\let\PYG@bf=\textbf\def\PYG@tc##1{\textcolor[rgb]{0.00,0.50,0.00}{##1}}}
\expandafter\def\csname PYG@tok@c\endcsname{\let\PYG@it=\textit\def\PYG@tc##1{\textcolor[rgb]{0.25,0.50,0.50}{##1}}}
\expandafter\def\csname PYG@tok@mf\endcsname{\def\PYG@tc##1{\textcolor[rgb]{0.40,0.40,0.40}{##1}}}
\expandafter\def\csname PYG@tok@err\endcsname{\def\PYG@bc##1{\setlength{\fboxsep}{0pt}\fcolorbox[rgb]{1.00,0.00,0.00}{1,1,1}{\strut ##1}}}
\expandafter\def\csname PYG@tok@mb\endcsname{\def\PYG@tc##1{\textcolor[rgb]{0.40,0.40,0.40}{##1}}}
\expandafter\def\csname PYG@tok@ss\endcsname{\def\PYG@tc##1{\textcolor[rgb]{0.10,0.09,0.49}{##1}}}
\expandafter\def\csname PYG@tok@sr\endcsname{\def\PYG@tc##1{\textcolor[rgb]{0.73,0.40,0.53}{##1}}}
\expandafter\def\csname PYG@tok@mo\endcsname{\def\PYG@tc##1{\textcolor[rgb]{0.40,0.40,0.40}{##1}}}
\expandafter\def\csname PYG@tok@kd\endcsname{\let\PYG@bf=\textbf\def\PYG@tc##1{\textcolor[rgb]{0.00,0.50,0.00}{##1}}}
\expandafter\def\csname PYG@tok@mi\endcsname{\def\PYG@tc##1{\textcolor[rgb]{0.40,0.40,0.40}{##1}}}
\expandafter\def\csname PYG@tok@kn\endcsname{\let\PYG@bf=\textbf\def\PYG@tc##1{\textcolor[rgb]{0.00,0.50,0.00}{##1}}}
\expandafter\def\csname PYG@tok@cpf\endcsname{\let\PYG@it=\textit\def\PYG@tc##1{\textcolor[rgb]{0.25,0.50,0.50}{##1}}}
\expandafter\def\csname PYG@tok@kr\endcsname{\let\PYG@bf=\textbf\def\PYG@tc##1{\textcolor[rgb]{0.00,0.50,0.00}{##1}}}
\expandafter\def\csname PYG@tok@s\endcsname{\def\PYG@tc##1{\textcolor[rgb]{0.73,0.13,0.13}{##1}}}
\expandafter\def\csname PYG@tok@kp\endcsname{\def\PYG@tc##1{\textcolor[rgb]{0.00,0.50,0.00}{##1}}}
\expandafter\def\csname PYG@tok@w\endcsname{\def\PYG@tc##1{\textcolor[rgb]{0.73,0.73,0.73}{##1}}}
\expandafter\def\csname PYG@tok@kt\endcsname{\def\PYG@tc##1{\textcolor[rgb]{0.69,0.00,0.25}{##1}}}
\expandafter\def\csname PYG@tok@sc\endcsname{\def\PYG@tc##1{\textcolor[rgb]{0.73,0.13,0.13}{##1}}}
\expandafter\def\csname PYG@tok@sb\endcsname{\def\PYG@tc##1{\textcolor[rgb]{0.73,0.13,0.13}{##1}}}
\expandafter\def\csname PYG@tok@k\endcsname{\let\PYG@bf=\textbf\def\PYG@tc##1{\textcolor[rgb]{0.00,0.50,0.00}{##1}}}
\expandafter\def\csname PYG@tok@se\endcsname{\let\PYG@bf=\textbf\def\PYG@tc##1{\textcolor[rgb]{0.73,0.40,0.13}{##1}}}
\expandafter\def\csname PYG@tok@sd\endcsname{\let\PYG@it=\textit\def\PYG@tc##1{\textcolor[rgb]{0.73,0.13,0.13}{##1}}}

\def\PYGZus{\char`\_}
\def\PYGZob{\char`\{}
\def\PYGZcb{\char`\}}

\def\PYGZam{\char`\&}

\def\PYGZdl{\char`\$}
\def\PYGZhy{\char`\-}
\def\PYGZsq{\char`\'}

\makeatother

\makeatletter
\def\PYGdefault@reset{\let\PYGdefault@it=\relax \let\PYGdefault@bf=\relax%
    \let\PYGdefault@ul=\relax \let\PYGdefault@tc=\relax%
    \let\PYGdefault@bc=\relax \let\PYGdefault@ff=\relax}
\def\PYGdefault@tok#1{\csname PYGdefault@tok@#1\endcsname}
\def\PYGdefault@toks#1+{\ifx\relax#1\empty\else%
    \PYGdefault@tok{#1}\expandafter\PYGdefault@toks\fi}
\def\PYGdefault@do#1{\PYGdefault@bc{\PYGdefault@tc{\PYGdefault@ul{%
    \PYGdefault@it{\PYGdefault@bf{\PYGdefault@ff{#1}}}}}}}
\def\PYGdefault#1#2{\PYGdefault@reset\PYGdefault@toks#1+\relax+\PYGdefault@do{#2}}

\expandafter\def\csname PYGdefault@tok@gd\endcsname{\def\PYGdefault@tc##1{\textcolor[rgb]{0.63,0.00,0.00}{##1}}}
\expandafter\def\csname PYGdefault@tok@gu\endcsname{\let\PYGdefault@bf=\textbf\def\PYGdefault@tc##1{\textcolor[rgb]{0.50,0.00,0.50}{##1}}}
\expandafter\def\csname PYGdefault@tok@gt\endcsname{\def\PYGdefault@tc##1{\textcolor[rgb]{0.00,0.27,0.87}{##1}}}
\expandafter\def\csname PYGdefault@tok@gs\endcsname{\let\PYGdefault@bf=\textbf}
\expandafter\def\csname PYGdefault@tok@gr\endcsname{\def\PYGdefault@tc##1{\textcolor[rgb]{1.00,0.00,0.00}{##1}}}
\expandafter\def\csname PYGdefault@tok@cm\endcsname{\let\PYGdefault@it=\textit\def\PYGdefault@tc##1{\textcolor[rgb]{0.25,0.50,0.50}{##1}}}
\expandafter\def\csname PYGdefault@tok@vg\endcsname{\def\PYGdefault@tc##1{\textcolor[rgb]{0.10,0.09,0.49}{##1}}}
\expandafter\def\csname PYGdefault@tok@vi\endcsname{\def\PYGdefault@tc##1{\textcolor[rgb]{0.10,0.09,0.49}{##1}}}
\expandafter\def\csname PYGdefault@tok@mh\endcsname{\def\PYGdefault@tc##1{\textcolor[rgb]{0.40,0.40,0.40}{##1}}}
\expandafter\def\csname PYGdefault@tok@cs\endcsname{\let\PYGdefault@it=\textit\def\PYGdefault@tc##1{\textcolor[rgb]{0.25,0.50,0.50}{##1}}}
\expandafter\def\csname PYGdefault@tok@ge\endcsname{\let\PYGdefault@it=\textit}
\expandafter\def\csname PYGdefault@tok@vc\endcsname{\def\PYGdefault@tc##1{\textcolor[rgb]{0.10,0.09,0.49}{##1}}}
\expandafter\def\csname PYGdefault@tok@il\endcsname{\def\PYGdefault@tc##1{\textcolor[rgb]{0.40,0.40,0.40}{##1}}}
\expandafter\def\csname PYGdefault@tok@go\endcsname{\def\PYGdefault@tc##1{\textcolor[rgb]{0.53,0.53,0.53}{##1}}}
\expandafter\def\csname PYGdefault@tok@cp\endcsname{\def\PYGdefault@tc##1{\textcolor[rgb]{0.74,0.48,0.00}{##1}}}
\expandafter\def\csname PYGdefault@tok@gi\endcsname{\def\PYGdefault@tc##1{\textcolor[rgb]{0.00,0.63,0.00}{##1}}}
\expandafter\def\csname PYGdefault@tok@gh\endcsname{\let\PYGdefault@bf=\textbf\def\PYGdefault@tc##1{\textcolor[rgb]{0.00,0.00,0.50}{##1}}}
\expandafter\def\csname PYGdefault@tok@ni\endcsname{\let\PYGdefault@bf=\textbf\def\PYGdefault@tc##1{\textcolor[rgb]{0.60,0.60,0.60}{##1}}}
\expandafter\def\csname PYGdefault@tok@nl\endcsname{\def\PYGdefault@tc##1{\textcolor[rgb]{0.63,0.63,0.00}{##1}}}
\expandafter\def\csname PYGdefault@tok@nn\endcsname{\let\PYGdefault@bf=\textbf\def\PYGdefault@tc##1{\textcolor[rgb]{0.00,0.00,1.00}{##1}}}
\expandafter\def\csname PYGdefault@tok@no\endcsname{\def\PYGdefault@tc##1{\textcolor[rgb]{0.53,0.00,0.00}{##1}}}
\expandafter\def\csname PYGdefault@tok@na\endcsname{\def\PYGdefault@tc##1{\textcolor[rgb]{0.49,0.56,0.16}{##1}}}
\expandafter\def\csname PYGdefault@tok@nb\endcsname{\def\PYGdefault@tc##1{\textcolor[rgb]{0.00,0.50,0.00}{##1}}}
\expandafter\def\csname PYGdefault@tok@nc\endcsname{\let\PYGdefault@bf=\textbf\def\PYGdefault@tc##1{\textcolor[rgb]{0.00,0.00,1.00}{##1}}}
\expandafter\def\csname PYGdefault@tok@nd\endcsname{\def\PYGdefault@tc##1{\textcolor[rgb]{0.67,0.13,1.00}{##1}}}
\expandafter\def\csname PYGdefault@tok@ne\endcsname{\let\PYGdefault@bf=\textbf\def\PYGdefault@tc##1{\textcolor[rgb]{0.82,0.25,0.23}{##1}}}
\expandafter\def\csname PYGdefault@tok@nf\endcsname{\def\PYGdefault@tc##1{\textcolor[rgb]{0.00,0.00,1.00}{##1}}}
\expandafter\def\csname PYGdefault@tok@si\endcsname{\let\PYGdefault@bf=\textbf\def\PYGdefault@tc##1{\textcolor[rgb]{0.73,0.40,0.53}{##1}}}
\expandafter\def\csname PYGdefault@tok@s2\endcsname{\def\PYGdefault@tc##1{\textcolor[rgb]{0.73,0.13,0.13}{##1}}}
\expandafter\def\csname PYGdefault@tok@nt\endcsname{\let\PYGdefault@bf=\textbf\def\PYGdefault@tc##1{\textcolor[rgb]{0.00,0.50,0.00}{##1}}}
\expandafter\def\csname PYGdefault@tok@nv\endcsname{\def\PYGdefault@tc##1{\textcolor[rgb]{0.10,0.09,0.49}{##1}}}
\expandafter\def\csname PYGdefault@tok@s1\endcsname{\def\PYGdefault@tc##1{\textcolor[rgb]{0.73,0.13,0.13}{##1}}}
\expandafter\def\csname PYGdefault@tok@ch\endcsname{\let\PYGdefault@it=\textit\def\PYGdefault@tc##1{\textcolor[rgb]{0.25,0.50,0.50}{##1}}}
\expandafter\def\csname PYGdefault@tok@m\endcsname{\def\PYGdefault@tc##1{\textcolor[rgb]{0.40,0.40,0.40}{##1}}}
\expandafter\def\csname PYGdefault@tok@gp\endcsname{\let\PYGdefault@bf=\textbf\def\PYGdefault@tc##1{\textcolor[rgb]{0.00,0.00,0.50}{##1}}}
\expandafter\def\csname PYGdefault@tok@sh\endcsname{\def\PYGdefault@tc##1{\textcolor[rgb]{0.73,0.13,0.13}{##1}}}
\expandafter\def\csname PYGdefault@tok@ow\endcsname{\let\PYGdefault@bf=\textbf\def\PYGdefault@tc##1{\textcolor[rgb]{0.67,0.13,1.00}{##1}}}
\expandafter\def\csname PYGdefault@tok@sx\endcsname{\def\PYGdefault@tc##1{\textcolor[rgb]{0.00,0.50,0.00}{##1}}}
\expandafter\def\csname PYGdefault@tok@bp\endcsname{\def\PYGdefault@tc##1{\textcolor[rgb]{0.00,0.50,0.00}{##1}}}
\expandafter\def\csname PYGdefault@tok@c1\endcsname{\let\PYGdefault@it=\textit\def\PYGdefault@tc##1{\textcolor[rgb]{0.25,0.50,0.50}{##1}}}
\expandafter\def\csname PYGdefault@tok@o\endcsname{\def\PYGdefault@tc##1{\textcolor[rgb]{0.40,0.40,0.40}{##1}}}
\expandafter\def\csname PYGdefault@tok@kc\endcsname{\let\PYGdefault@bf=\textbf\def\PYGdefault@tc##1{\textcolor[rgb]{0.00,0.50,0.00}{##1}}}
\expandafter\def\csname PYGdefault@tok@c\endcsname{\let\PYGdefault@it=\textit\def\PYGdefault@tc##1{\textcolor[rgb]{0.25,0.50,0.50}{##1}}}
\expandafter\def\csname PYGdefault@tok@mf\endcsname{\def\PYGdefault@tc##1{\textcolor[rgb]{0.40,0.40,0.40}{##1}}}
\expandafter\def\csname PYGdefault@tok@err\endcsname{\def\PYGdefault@bc##1{\setlength{\fboxsep}{0pt}\fcolorbox[rgb]{1.00,0.00,0.00}{1,1,1}{\strut ##1}}}
\expandafter\def\csname PYGdefault@tok@mb\endcsname{\def\PYGdefault@tc##1{\textcolor[rgb]{0.40,0.40,0.40}{##1}}}
\expandafter\def\csname PYGdefault@tok@ss\endcsname{\def\PYGdefault@tc##1{\textcolor[rgb]{0.10,0.09,0.49}{##1}}}
\expandafter\def\csname PYGdefault@tok@sr\endcsname{\def\PYGdefault@tc##1{\textcolor[rgb]{0.73,0.40,0.53}{##1}}}
\expandafter\def\csname PYGdefault@tok@mo\endcsname{\def\PYGdefault@tc##1{\textcolor[rgb]{0.40,0.40,0.40}{##1}}}
\expandafter\def\csname PYGdefault@tok@kd\endcsname{\let\PYGdefault@bf=\textbf\def\PYGdefault@tc##1{\textcolor[rgb]{0.00,0.50,0.00}{##1}}}
\expandafter\def\csname PYGdefault@tok@mi\endcsname{\def\PYGdefault@tc##1{\textcolor[rgb]{0.40,0.40,0.40}{##1}}}
\expandafter\def\csname PYGdefault@tok@kn\endcsname{\let\PYGdefault@bf=\textbf\def\PYGdefault@tc##1{\textcolor[rgb]{0.00,0.50,0.00}{##1}}}
\expandafter\def\csname PYGdefault@tok@cpf\endcsname{\let\PYGdefault@it=\textit\def\PYGdefault@tc##1{\textcolor[rgb]{0.25,0.50,0.50}{##1}}}
\expandafter\def\csname PYGdefault@tok@kr\endcsname{\let\PYGdefault@bf=\textbf\def\PYGdefault@tc##1{\textcolor[rgb]{0.00,0.50,0.00}{##1}}}
\expandafter\def\csname PYGdefault@tok@s\endcsname{\def\PYGdefault@tc##1{\textcolor[rgb]{0.73,0.13,0.13}{##1}}}
\expandafter\def\csname PYGdefault@tok@kp\endcsname{\def\PYGdefault@tc##1{\textcolor[rgb]{0.00,0.50,0.00}{##1}}}
\expandafter\def\csname PYGdefault@tok@w\endcsname{\def\PYGdefault@tc##1{\textcolor[rgb]{0.73,0.73,0.73}{##1}}}
\expandafter\def\csname PYGdefault@tok@kt\endcsname{\def\PYGdefault@tc##1{\textcolor[rgb]{0.69,0.00,0.25}{##1}}}
\expandafter\def\csname PYGdefault@tok@sc\endcsname{\def\PYGdefault@tc##1{\textcolor[rgb]{0.73,0.13,0.13}{##1}}}
\expandafter\def\csname PYGdefault@tok@sb\endcsname{\def\PYGdefault@tc##1{\textcolor[rgb]{0.73,0.13,0.13}{##1}}}
\expandafter\def\csname PYGdefault@tok@k\endcsname{\let\PYGdefault@bf=\textbf\def\PYGdefault@tc##1{\textcolor[rgb]{0.00,0.50,0.00}{##1}}}
\expandafter\def\csname PYGdefault@tok@se\endcsname{\let\PYGdefault@bf=\textbf\def\PYGdefault@tc##1{\textcolor[rgb]{0.73,0.40,0.13}{##1}}}
\expandafter\def\csname PYGdefault@tok@sd\endcsname{\let\PYGdefault@it=\textit\def\PYGdefault@tc##1{\textcolor[rgb]{0.73,0.13,0.13}{##1}}}

\makeatother

\begin{document}

\title{Yadage and Packtivity -- analysis preservation using parametrized workflows}

\author{Kyle Cranmer$^1$ and Lukas Heinrich$^1$}
\address{$^1$ Department of Physics, New York University, New York, USA}

\ead{lukas.heinrich@cern.ch}

\begin{abstract}
	Preserving data analyses produced by the collaborations at LHC in a parametrized fashion is crucial in order to maintain reproducibility and re-usability. We argue for a declarative description in terms of individual processing steps -- ``packtivities'' -- linked through a dynamic directed acyclic graph (DAG) and present an initial set of JSON schemas for such a description and an implementation -- ``yadage'' --  capable of executing workflows of analysis preserved via Linux containers.
\end{abstract}

\section{Introduction}

Data analyses of LHC data consist of workflows that utilize a diverse set of software tools to produce physics results. The tools range from large software frameworks like Gaudi\cite{Gaudi2001} to single-purpose scripts written by individual analyzers or analysis teams. The analysis steps that lead to a particular physics result are often not reproducible without significant assistance from the original authors. This severely limits the capability to re-execute the original analysis or to re-use its analysis procedures in new contexts. An important application for such re-use is the systematic re-interpretation of a given analysis with respect to alternative models of new physics\cite{Cranmer2011}. Therefore, it is desirable to have a system to archive analysis code as well as the analysis procedure in a manner, that enables both re-execution and re-use. This document presents work on workflow capture that addresses these issues in a platform and language-agnostic manner.

\subsection{Short anatomy of analysis workflows}

The driving paradigm of LHC analyses is the selection of events within the experiments' dataset and, typically, comparing those events to expectations derived using both data-driven techniques and Monte-Carlo simulations. Since every collision event (whether real or simulated) is independent of the others, the data analysis problem becomes \emph{embarrassingly parallel}. Consequently, the most common task in a LHC analysis is the parallel processing of events by algorithms that transform the event data into higher-level representations (e.g. from raw detector data to reconstructed `analysis objects`) or perform event selection or otherwise reduce the dataset size, for example by selectively storing only partial event information (`thinning').

The main reconstruction transformations are often handled either on a collaboration-wide or physics working group level and use centrally managed and documented code with fixed release schedules and procedures. Transform configurations, such as the used executable and its command line options, are managed centrally as well (e.g. with databases such as AMI\cite{AMI2010}). Therefore, these operations are comparatively easy to preserve and reproduce.

On the other hand, custom code developed by the end-user analysis team is often much harder to reproduce due to the diversity of tools, workflows and computing environments that are used by an individual analysis team. In the case of the ATLAS experiment, a very wide spectrum of analysis frameworks have been used during Run-1 to analyze events in the ``post-AOD'' stage, i.e. after central reconstruction. This ranged from large and complex frameworks such as \texttt{SFrame} or \texttt{Athena}, handling not only the main event-loop, but also managing calibration tool instantiation and data-handling, to pure ROOT-based programs such as \texttt{TTree::MakeClass}- and \texttt{TTree::MakeSelector}-based codes. Since Run-2, ATLAS has seen a increasing level of homogenization in analysis codes, where many groups use one of two high-level analysis frameworks within which they develop the custom routines needed for the analysis at hand.

Once all data (real and simulated) is sufficiently reduced, usually a statistical analysis is performed, in which the observed data is compared to the expectations given by the physics model under study. Here, a range of statistics packages such as HistFitter and HistFactory or loosely-structured scripts, that utilize RooFit/RooStats directly, is used to extract the relevant physics results such as interval estimates on model parameters. This can include precision measurements of Standard Model observables or exclusion limits on parameters of models of physics beyond the Standard Model.

\subsection{Analysis preservation for re-use}
In the context of analysis preservation, the entire analysis can viewed as an abstract function that maps data and the model hypothesis to the analysis results:

\begin{equation}
	\textrm{result} = f_{\textrm{analysis}}(\textrm{data}, \textrm{model})
\end{equation}

Ideally, one would like to preserve this map in a completely parametrized form, $f_\mathrm{analysis}(\cdot,\cdot)$ independent of the specific data and model on which it has been applied to obtain the result at hand. Realistically, however, the analysis is tightly coupled to the recorded data it was developed against, due to various reasons such as file formats and re-processing versions. The model-dependence, on the other hand, can often times be factored out more easily, especially for analyses that search for physics phenomena beyond the Standard Model, where the Beyond the Standard Model (BSM) contribution is estimated separately from the Standard Model backgrounds. An analysis preservation approach that is designed to be model-independent would thus enable both re-interpretation and statistical combinations of multiple analyses after the initial publication.

In order for such to achieve such a parametrized preservation, two separate types of information need to be captured:

\begin{enumerate}
	\item a descriptions of the individual parametrized analysis steps such as event selection steps or the subsequent statistical analysis
	\item a description of the workflow that logically links these individual steps in order to arrive at the analysis result data
\end{enumerate}

In this document we introduce schemas to capture this information in flat JSON data as well as a framework to read back that information and re-execute such a preserved analysis.

\section{Capturing parametrized activities}

An appropriate model to capture the different steps of an analysis is the data model employed by the W3C PROV standard\cite{w3c-prov-dm}, in which the basic ingredients are \emph{entities} and \emph{activities} to track data provenance. Activities act on existing entities and generate new ones. In the context of an HEP analysis, an entity is often a set of files (such as a dataset) or a data product derived from them, while an activity is most often the execution of a piece of software that takes entities (i.e. data) as input and generates new entities as outputs, e.g. by writing a new set of files.  These operations can be parametrized by a few variables such that the activity appears as a function of the parameters $p_1,p_2,\cdots,p_n$ and some notion of an input state $\sigma$ (this could for example be a filesystem directory), which may be modified as a side-effect of the function.

\begin{equation}
\mathrm{output} = f_{\mathrm{step}}(p_1,p_2,\cdots,p_n, \sigma),
\label{eq:singlestep}
\end{equation}

It is useful to partition the return value of this function into a tuple of an output state after processing $\sigma'$ and a separate record of human- and machine-readable \emph{result data}, $\mathbf{r}$, that provides additional machine-readable data, possibly describing the side effects such as filesystem paths of files generated during this step. This separation allows for a convenient definition of workflows later on, as each step identifies and publishes the relevant data fragments (i.e. entities) it produces.

\begin{equation}
(\mathbf{r},\sigma') = f_{\mathrm{step}}(\mathbf{p}, \sigma),
\label{eq:singlestep_v2}
\end{equation}

An activity is thus an abstract interface that transforms parameters into result data while modifying an external state. As an interchange format for both the input parameters $\mathbf{p}$ and result data $\mathbf{r}$ JSON is a suitable choice.

The information required to fully capture such parametrized activities may be partitioned into three basic pieces:

\begin{description}
    \item[process] a parametrized description with which one can produce a fully-defined activity description -- the (``job'') -- based on concrete parameters (such as a templated command line)
    \item[\emph{environment}] a description of the environment in which this job is to be executed. This may for example include a description of the necessary software to run the above process
    \item[publisher] a description of how to extract the relevant information or data fragments subsequent to the execution of the job
\end{description}.

For each of these pieces, multiple concrete implementations are possible. Irrespective of the implementation, the basic procedure for execution (given some execution backend) is shown in algorithm~\ref{alg:packtivity}.
\\

\begin{algorithm}[H]
\footnotesize
 \KwIn{$\mathbf{p}$, $\sigma$}
 \KwOut{$\mathbf{r}$, $\sigma'$}
 \SetKwFunction{Proc}{Process}
 \SetKwFunction{Bknd}{Backend}
 \SetKwFunction{Pblsh}{Publisher}
 \Begin{
   job $\leftarrow$  \Proc{$\mathbf{p}$}\;
   $\sigma' \leftarrow$  \Bknd{$\sigma'$,job,environment}\;
   $\mathbf{r} \leftarrow$  \Pblsh{job,$\sigma'$}\;
   \Return{$\mathbf{r}$},$\sigma'$
}
 \caption{Activity($\mathbf{p}$,$\sigma$)}
 \label{alg:packtivity}
\end{algorithm}

\subsection{Packtivity}

To capture such `packaged activities` -- or `packtivities' -- a extensible set of JSON schemas have been developed to describe the three interfaces -- process, environment, publisher -- identified in the previous section. The choice is motivated by the simplicity and ubiquity of the JSON format, which makes it suitable for long-term and implementation-independent archival. Sub-schemas will generally be identified by a interface-wide `property` but implementation specific `property value`. An examples of full packtivity definitions are provided in listing~\ref{mnt:fullpacktivity}. A number of JSON schemas are collected under the \verb+yadage-schemas+ package available at GitHub and via PyPI\cite{yadage-schemas}

\begin{listing}[!ht]
{\footnotesize
\begin{Verbatim}[commandchars=\\\{\}]
\PYG{l+lScalar+lScalarPlain}{process}\PYG{p+pIndicator}{:}
  \PYG{l+lScalar+lScalarPlain}{process\PYGZus{}type}\PYG{p+pIndicator}{:} \PYG{l+s}{\PYGZsq{}string\PYGZhy{}interpolated\PYGZhy{}cmd\PYGZsq{}}
  \PYG{l+lScalar+lScalarPlain}{cmd}\PYG{p+pIndicator}{:} \PYG{l+s}{\PYGZsq{}DelphesHepMC}\PYG{n+nv}{  }\PYG{l+s}{\PYGZob{}delphes\PYGZus{}card\PYGZcb{}}\PYG{n+nv}{ }\PYG{l+s}{\PYGZob{}outputroot\PYGZcb{}}\PYG{n+nv}{ }\PYG{l+s}{\PYGZob{}inputhepmc\PYGZcb{}}\PYG{n+nv}{ }\PYG{l+s}{\PYGZam{}\PYGZam{}}\PYG{n+nv}{ }\PYG{l+s}{root2lhco}\PYG{n+nv}{ }\PYG{l+s}{\PYGZob{}outputroot\PYGZcb{}}\PYG{n+nv}{ }\PYG{l+s}{\PYGZob{}outputlhco\PYGZcb{}\PYGZsq{}}
\PYG{l+lScalar+lScalarPlain}{publisher}\PYG{p+pIndicator}{:}
  \PYG{l+lScalar+lScalarPlain}{publisher\PYGZus{}type}\PYG{p+pIndicator}{:} \PYG{l+s}{\PYGZsq{}frompar\PYGZhy{}pub\PYGZsq{}}
  \PYG{l+lScalar+lScalarPlain}{outputmap}\PYG{p+pIndicator}{:}
    \PYG{l+lScalar+lScalarPlain}{lhcofile}\PYG{p+pIndicator}{:} \PYG{l+lScalar+lScalarPlain}{outputlhco}
    \PYG{l+lScalar+lScalarPlain}{rootfile}\PYG{p+pIndicator}{:} \PYG{l+lScalar+lScalarPlain}{outputroot}
\PYG{l+lScalar+lScalarPlain}{environment}\PYG{p+pIndicator}{:}
  \PYG{l+lScalar+lScalarPlain}{environment\PYGZus{}type}\PYG{p+pIndicator}{:} \PYG{l+s}{\PYGZsq{}docker\PYGZhy{}encapsulated\PYGZsq{}}
  \PYG{l+lScalar+lScalarPlain}{image}\PYG{p+pIndicator}{:} \PYG{l+lScalar+lScalarPlain}{lukasheinrich/root\PYGZhy{}delphes}
\end{Verbatim}
}
\caption{An example packtivity manifest}
\label{mnt:fullpacktivity}
\end{listing}

\subsubsection{Process descriptions}
A form of capturing parametrized process information, that is accessible and convenient for analysis teams to produce, are templated string of multi-line scripts or single-line command lines from which concrete job manifests are formed by interpolating these strings based on input parameters provided as JSON documents. In the example shown in listing~\ref{mnt:fullpacktivity}, the process has five replacement fields that need to be provided by the input JSON document, for example by having a top-level object with appropriately named properties.

\subsubsection{Environment descriptions}
Describing the software environment or run-time that the job formed by the `process' requires is a trade-off between completeness and convenience and could range from specifying merely a specific software release number to a full virtual machine image that includes both hardware and software virtualization. In practice, using Linux container technologies such as Docker have proven to be a useful middle-ground.

In HEP contexts, a large amount of software is installed centrally in the global, read-only filesystem CVMFS, thus it may not be feasible for all application to provide a standalone CVMFS-independent installation. However, as CVMFS is a versioned filesystem, in principle it is possible to mount it at the version that the original analysis executed against. Similarly, some applications may require additional run-time data such as secrets used for VO authentication. Generally, the number of such external dependencies should be kept at a minimum for preservation purposes.

\subsubsection{Publisher descriptions}
The execution of the process in a given environment typically results in the modification of an external state which may, for example, be provided by mounting an external shared filesystem such as CephFS into the containers. However in order to describe multi-step analysis workflows, it is necessary to have a semantic description of what the relevant data fragments of this activity are. As laid out in the previous section, JSON is a suitable format. The processes themselves do not necessarily produce JSON data, so that it is helpful to include the notion of an external `publishing manifest' into the packtivity definition, that a implementation can use in order to to derive a JSON object. 

Often, the result JSON data can be derived by simple inspection of the input parameters such as in listing~\ref{mnt:fullpacktivity}. In this case it suffices to provide a mapping from input parameters to output keys. In other scenarios the result data may only be fully formed after the process execution, for example if the process generates a dynamic number of files that all need to be published. In this case other publishers may be defined, such as using glob patterns or regular expressions.

\subsection{Reference implementation}

The python-based \verb+packtivity+\cite{packtivity} package provides an implementation of algorithm~\ref{alg:packtivity} for a number of different backends. Both synchronous and asynchronous backends such as Celery or IPython Clusters are provided for the currently defined process-types and environments. Both python language bindings as well as command line executables are provided to run packtivities.

\section{Parametrized workflow model}

The packaging of the individual processing activities captures a lot of the information needed to re-produce or re-use a given analysis. The command line interface or the language bindings can be used in code to execute pre-packaged activities in a suitable order. It is, however, desirable to capture this workflow logic in a similarly declarative fashion as the packtivities themselves, such that its execution may be automated.

A suitable data model for the description of workflows is the directed acyclic graph (DAG). In such a graph, nodes represent individual activities, while directed edges denote dependency relations of activities. This allows to capture non-linear workflows and enables the distribution and orchestration of analysis workflows across distributed systems.

As noted, for re-use applications, it is important that the workflows are parametrized. This may in turn introduce some parameter-dependency on not only the parameters for individual packtivities, but also the topology of the workflow graph. A number of DAG-based workflow systems such as DAGMan or the Common Workflow Language \cite{HTCondorDAGMan,Amstutz2016} exist, however those tools introduce limitations that hinder the definition of workflow in those parameter-dependent scenarios. Therefore, a number an extensible workflow definition system has been developed that grants first-class status to parametrized DAGs.

\subsection{Workflow stages}

The central tenet of parametrized and dynamic workflows is that, instead of archiving DAG descriptions that fully fix the topology, one should rather store sufficient information into a \emph{workflow template} $T$, from which it is possible construct these DAGs -- \emph{workflow instances} $W$ -- during run-time once sufficient information (such as parameter values) is available. As such, the workflow template T is made up from a set of \emph{workflow stages} $s_i$ and $T = \{s_1,s_2\dots s_n\}$. A stage represents a piece logic that add nodes and edges to the instance DAG.

The time at which this operation may be applied can be dependent on the state of the instance. For example it may require that some node within the instance graph has already been processed by a backend and its result data is known. Therefore a stage is defined by two pieces of information

\begin{itemize}
	\item a definition of dependencies which the workflow instance $W$ must fulfill for the stage to be applicable. This is conveniently expressed as a DAG-valued \emph{predicate function} $d: W\to \{\mathrm{true},\mathrm{false}\}$
	\item a DAG-valued function that, given a workflow instance, returns an updated workflow instance with additional nodes and edges or newly defined stages. This may be expressed as a DAG-valued \emph{scheduler function} $f: W\to W'$
\end{itemize}

In order to process a parametrized workflow defined by such a template, a simple algorithm, shown in algorithm~\ref{alg:adage} may be followed in order to continuously monitor a workflow instance. The algorithm applies the stages' scheduler functions as soon as its predicate is fulfilled and submit individual nodes to a packtivity backend.
\\

\begin{algorithm}[H]
\footnotesize
 \KwIn{Workflow Template, Initialization JSON}
 \KwOut{Complete set of Workflow data entities}
 initialize new workflow instance (empty DAG) wflow;

 \While{there exist unapplied stages}{
  \ForEach{stage in unapplied stages}{
    \If{stage applicable (predicate returns True)}{
      expand DAG with stage as per scheduler function
    }
  }
  \ForEach{node in DAG}{
    \If{node has not been submitted to backend and all incoming edges succeeded}{
      submit node to backend
    }
  }

 }
 wait until all nodes processed by backend
 \caption{Basic Yadage Workflow Engine}
 \label{alg:adage}
\end{algorithm}

\subsection{Yadage}

\subsubsection{Workflow Definition}

As for packtivities, the \verb+yadage-schemas+ package includes JSON schemas to define workflow stages. From experience complex workflows may be defined using a small number of stage-types. Currently stages that schedule one or more nodes of a singled packtivity (each with different parameters) are defined. In the case of multi-node stages, a number of scattering patterns maybe used. An example stage definition is shown in listing~\ref{mnt:stage}. The dependencies are listed by naming other stages. The current stage is then considered applicable if all nodes by the dependent stage are successfully processed. The scheduler function $f$, is defined under the \verb+scheduler+ property and includes instructions on how to access the result data of dependent nodes added to the Graph by dependent stages in order to define the parameters of the packtivity to be scheduled. As seen in the listing, JSON References are used in order to reference packtivity definitions via URIs.

\begin{listing}[!ht]
{\footnotesize
\begin{Verbatim}[commandchars=\\\{\}]
\PYG{l+lScalar+lScalarPlain}{name}\PYG{p+pIndicator}{:} \PYG{l+lScalar+lScalarPlain}{delphes}
\PYG{l+lScalar+lScalarPlain}{dependencies}\PYG{p+pIndicator}{:} \PYG{p+pIndicator}{[}\PYG{n+nv}{pythia}\PYG{p+pIndicator}{]}
\PYG{l+lScalar+lScalarPlain}{scheduler}\PYG{p+pIndicator}{:}
  \PYG{l+lScalar+lScalarPlain}{scheduler\PYGZus{}type}\PYG{p+pIndicator}{:} \PYG{l+s}{\PYGZsq{}singlestep\PYGZhy{}stage\PYGZsq{}}
  \PYG{l+lScalar+lScalarPlain}{step}\PYG{p+pIndicator}{:} \PYG{p+pIndicator}{\PYGZob{}}\PYG{n+nv}{\PYGZdl{}ref}\PYG{p+pIndicator}{:} \PYG{l+s}{\PYGZsq{}delphes.yml\PYGZsq{}}\PYG{p+pIndicator}{\PYGZcb{}}
  \PYG{l+lScalar+lScalarPlain}{parameters}\PYG{p+pIndicator}{:}
    \PYG{l+lScalar+lScalarPlain}{outputroot}\PYG{p+pIndicator}{:} \PYG{l+s}{\PYGZsq{}\PYGZob{}workdir\PYGZcb{}/output.root\PYGZsq{}}
    \PYG{l+lScalar+lScalarPlain}{outputlhco}\PYG{p+pIndicator}{:} \PYG{l+s}{\PYGZsq{}\PYGZob{}workdir\PYGZcb{}/output.lhco\PYGZsq{}}
    \PYG{l+lScalar+lScalarPlain}{delphes\PYGZus{}card}\PYG{p+pIndicator}{:} \PYG{l+s}{\PYGZsq{}delphes/cards/delphes\PYGZus{}card\PYGZus{}ATLAS.tcl\PYGZsq{}}
    \PYG{l+lScalar+lScalarPlain}{inputhepmc}\PYG{p+pIndicator}{:} \PYG{p+pIndicator}{\PYGZob{}}\PYG{n+nv}{stages}\PYG{p+pIndicator}{:} \PYG{n+nv}{pythia}\PYG{p+pIndicator}{,} \PYG{n+nv}{output}\PYG{p+pIndicator}{:} \PYG{n+nv}{hepmcfile}\PYG{p+pIndicator}{\PYGZcb{}}
\end{Verbatim}
}
\caption{An Example Yadage Stage Manifest}
\label{mnt:stage}
\end{listing}

\subsection{Workflow composition}

An important feature of workflows defined via the yadage schemas is composability that is not dependent on coordination between workflow authors. HEP workflows can involve many different stages in order to transform real or simulated events from detector or even particle-level data all the way to a final analysis result. Parts of these workflows may be primarily defined by different groups. For example, the workflow to describe the generation of Monte Carlo events based on a certain model may be defined primarily by physics working groups, while the reconstruction chain to transform generated events into fully reconstructed events (`AOD' data) usually is the responsibility of a central reconstruction group. Finally the downstream analysis of those reconstructed events is done on the analysis-team level. Each of these macro-parts of the workflow may be a multi-stage workflow themselves, such that the ability to compose them without modification into a larger workflow is desirable. In the case of the currently defined yadage stages, this is easily achieved by modifying the stage schedule 'workflows' instead of individual packtivities. In this case, the scheduling function does not add nodes or edges, but rather adds newly defined stages corresponding to the sub-workflows to the instances list of stages, that subsequently will be applied in a scoped fashion to exclude the possibility of e.g. naming collisions/ambiguities.

\subsubsection{Reference implementation}

Workflows declared using the above schemas may be executed using the \verb+yadage+ package\cite{yadage} that implements the basic scheduling and submission algorithm outlined above and transparently is able to use any packtivity backend implementation. Further, it implements a number of convenience features such as caching/memoization of individual packtivity results and visualization capabilities.  

\section{Applications}

\subsection{Run-I reinterpretation}
Workflows defined via the schemas outlined above have been used in a number of different contexts. ATLAS has published multiple re-interpretations of analyses prepared in Run-1. In these campaigns, a number of analyses designed to investigate particular supersymmetric scenarios have been re-interpreted to derive a more comprehensive assessment of the ATLAS experiments' sensitivity to supersymmetry such as under the 19-dimensional `phenomenological MSSM' (pMSSM) or a more restricted five-dimensional scan targeting electroweak sparticle production\cite{Aad:2015baa,Aaboud:2016wna,ATLAS-CONF-2016-033}.

\subsection{CERN Analysis Preservation Portal}

The CERN Analysis Preservation Portal (CAP)\cite{Chen2016} aims to preserve analysis information in the form of a digital library. Besides assembling metadata of analyses, such as the involved researchers and institutes, it also seeks to archive more technical information, such as code repositories and software environments. The workflow and activity schemas described here have been deeply integrated into the CAP system. Thanks to the choice of JSON schemas, the can be treated as native data in the context of the Invenio Digitial Library framework\cite{Invenio} facilitating, for example, discoverability and composability of workflow pieces. 

\subsection{RECAST}

As noted, LHC experiments already engage in re-interpretation campaigns. However, the current approach requires a high level of coordination between analysis groups for each re-interpretation. RECAST is a framework in which the re-interpretation is streamlined by utilizing workflows that are archived in a re-usable manner. Originally proposed in 2010, a prototype backend implementation has been recently been developed and deployed at CERN. RECAST allows interested parties outside of the LHC collaboration to suggest for re-interpretation, by providing model information such as parameter cards in the SLHA format to the experiments using a web-based interface or, alternatively, REST and Python APIs. Upon review by the experiments, the experiment may decide to re-run an archived analysis based on these new model inputs and provide a response in the form of likelihood information (e.g. $CL_s$ values)

The prototype backend consists of a `control-center' web-service that is accessible using VO-filtered CERN Single-Sign-On. This web-service displays incoming requests and allows operators to launch a re-execution of an analysis based on the RECAST request. The actual workflow execution is then handed off to a distributed system. Here, yadage and packtivity are heavily used to both define workflows and drive their execution. In the course of this development, a packtivity backend has been implemented that allows the scheduling of HEP container workloads on a Kubernetes Cluster deployed on the CERN OpenStack infrastructure using OpenStack Magnum. The integration with cloud-native tools allows for convenient scaling characteristics and a highly distributed workflow execution that may be monitored in real-time using the web-interface of the `control-center'.

\section{Summary}

We have presented a framework to define parametrized workflows in order to preserve high-energy physics analyses in a format that allows collaboration members to re-execute the original analysis in the context of new physics models. The framework defines a set of portable JSON schemas that describe both the individual processing steps and and workflow logic to orchestrate multiple steps. We also presented language-agnostic algorithms to re-execute analyses based on these descriptions. Furthermore, reference implementations of these algorithms distributed as python-based packages (\verb+packtivity+ and \verb+yadage+) were presented. Workflows definitions and the reference implementation have been used in the past for re-interpretation campaigns within the ATLAS collaboration and are deeply integrated in the CERN Analysis Preservation Portal and RECAST projects. 

\section{Acknowledgements}

Cranmer and Heinrich are both supported through NSF ACI-1450310, additionally Cranmer is supported by PHY-1505463.

\section*{References}
\bibliography{chep2016_yadage}
\bibliographystyle{iopart-num}

\end{document}